\documentclass[letterpaper, 10 pt, journal, twoside]{ieeetran}
\usepackage{cite}
\usepackage{amsmath,amssymb,amsfonts}
\usepackage{algorithmic}
\usepackage{graphicx}
\usepackage{subfigure}
\usepackage{xcolor}

\ifCLASSINFOpdf
\else
\fi

\begin{document}

\newpage
\onecolumn

\thispagestyle{empty}

\begin{centering}

\empty


\vspace{5cm}

 Please see the most recent version in the IEEE version.

\vspace{5cm}

 DOI: 10.1109/LRA.2024.3360815

 IEEE Xplore: https://ieeexplore.ieee.org/document/10417076

\vspace{8cm} 
\centering A. Adeyemi, U. Sen, S. M. Ercan and M. Sarac, "Hand Dominance and Congruence for Wrist-Worn \\ Haptics Using Custom Voice-Coil Actuation," in \textit{IEEE Robotics and Automation} Letters, vol. 9, no. 4, \\ pp. 3053-3059, April 2024, doi: 10.1109/LRA.2024.3360815.

\vspace{5cm} 
© 2024 IEEE.  Personal use of this material is permitted.  Permission from IEEE must be obtained for all other uses, in any current or future media, including reprinting/republishing this material for advertising or promotional purposes, creating new collective works, for resale or redistribution to servers or lists, or reuse of any copyrighted component of this work in other works.
 \strut
\end{centering}


\twocolumn

\title{\LARGE \bf Hand Dominance and Congruence for Wrist-worn Haptics using Custom Voice-Coil Actuation }


\author{
Ayoade Adeyemi,
Umit Sen, 
Samet Mert Ercan, 
Mine Sarac$^{*}$, ~\IEEEmembership{Member,~IEEE}
\thanks{This work is funded by TUBİTAK 
2232-B International Fellowship for Early Stage Researchers Program, number 121C147.}
\thanks{$^{*}$Kadir Has University, \.Istanbul, Turkey. \\
 {\tt\footnotesize ayoade.adeyemi,umit.sen,sametmert@stu.khas.edu.tr} \\
{\tt\footnotesize mine.sarac@khas.edu.tr}}
}

\markboth{}%
{}
%




\IEEEtitleabstractindextext{%
\begin{abstract}
During virtual interactions, rendering haptic feedback on a remote location (like the wrist) instead of the fingertips freeing users' hands from mechanical devices. This allows for real interactions while still providing information regarding the mechanical properties of virtual objects. In this paper, we present CoWrHap -- a novel wrist-worn haptic device with custom-made voice coil actuation to render force feedback. Then, we investigate the impact of asking participants to use their dominant or non-dominant hand for virtual interactions and the best mapping between the active hand and the wrist receiving the haptic feedback, which can be defined as hand-wrist congruence through a user experiment based on a stiffness discrimination task. Our results show that participants performed the tasks \textit{(i)} better with non-congruent mapping but reported better experiences with congruent mapping, and \textit{(ii)} with no statistical difference in terms of hand dominance but reported better user experience and enjoyment using their dominant hands. This study indicates that participants can perceive mechanical properties via haptic feedback provided through CoWrHap. 
\end{abstract}

}

\maketitle
 \pagenumbering{arabic}

\IEEEdisplaynontitleabstractindextext

%
\IEEEpeerreviewmaketitle


\section{Introduction}
%
%
%
%


Haptics is crucial to improve the quality of user performance and experience during interactions in Virtual and Augmented Reality (VR/AR) environments. Rendering haptic feedback based on the physical properties of a virtual object is useful for information transfer (e.g., medical simulators~\cite{Sarac2021_v1}) or overall task performance (e.g., exploration tasks~\cite{gaffary_ar_2017}). Further research introducing haptic feedback to VR scenarios indicates a significant increase in task performance~\cite{Swapp:2006} or immersion~\cite{Salln:2000}.


Most haptic devices are designed as wearable, tactile devices for the fingertips \cite{maisto_evaluation_2017, bortone_wearable_2018} due to the highest mechanoreceptor intensity at the fingertips \cite{Johansson1979} and the most intuitive and realistic sensation for users. However, rendering haptic feedback on the fingertips comes with drawbacks too. To increase comfort and wearability, they must be designed in small sizes; thus must be equipped with powerful but minimized actuators -- increasing the overall cost. Even with the smallest design possible, they might limit the transparency of natural hand movements. Finally, they might prevent the interaction capabilities with virtual objects -- especially during AR interactions.


Relocating the haptic feedback from the fingertips to an alternative body location (like the wrist and foot \cite{fukuda2018visual}) can address these issues and challenges by freeing the hands and still rendering useful (and believable) information regarding virtual interactions. 
The limitations of mechanoreceptor density might be overcome by choosing more advanced actuators and sensors for wrist-worn haptic devices. The literature already has many examples of wrist-worn haptic devices designed for information transfer scenarios \cite{paredes_synestouch_2015, alarcon_design_2017, gaudeni_presenting_2019, maeda_hapticaid_2016} or VR/AR interaction scenarios \cite{Pezent2022_wrist, Moriyama2022_forearm, Sarac2022}.

\begin{figure}[t!]%
\centering
\includegraphics[width=0.48\textwidth]{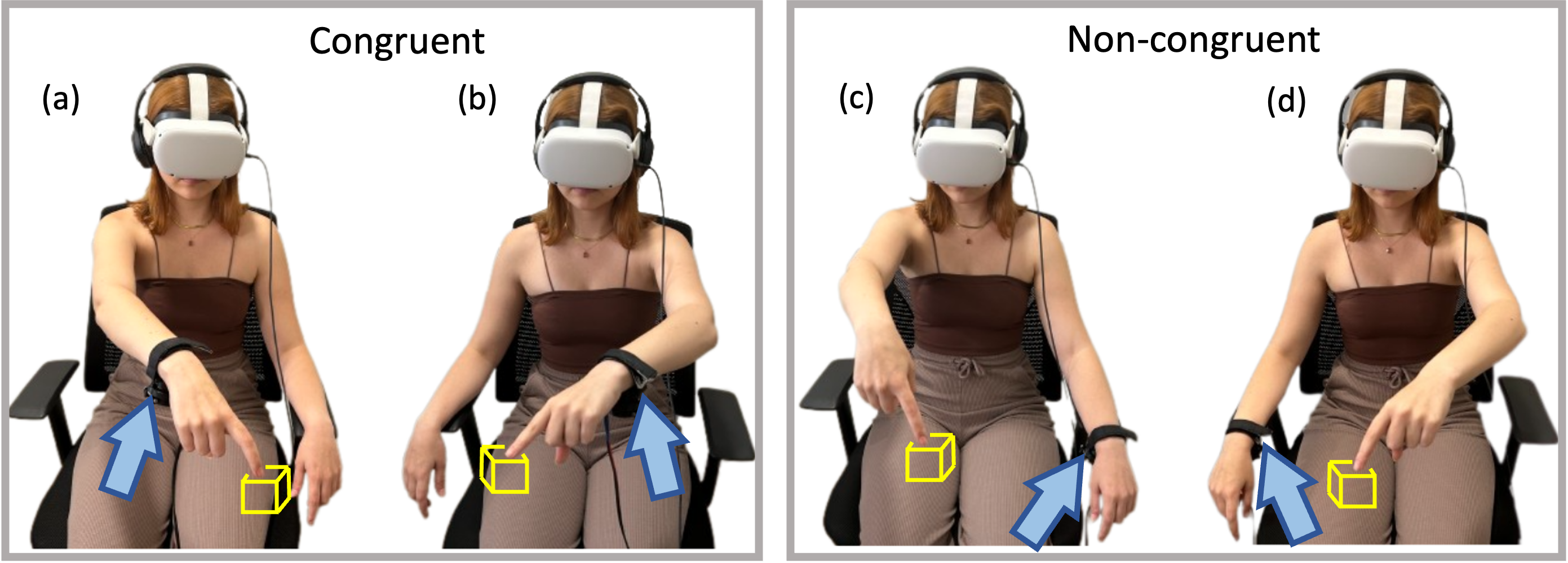}
 \vspace*{-0.5\baselineskip}
 \caption{Experiment Conditions: The participant receives either congruent haptics (on the same side as the dominant or non-dominant hand interacting with VR (a-b)) or non-congruent haptics (on the opposite side as the dominant or non-dominant hand interacting with VR (c-d)). The yellow box represents the virtual object, while the blue arrows indicate the haptic location.}
 \vspace*{-1.2\baselineskip}
 \label{fig:conditions}%
\end{figure}


In addition, 
how to relocate the haptic feedback from the fingertip to the wrist is an important question that needs to be addressed in many alternative mapping scenarios. Previous studies investigated the user performance as participants explored mechanical properties for different sides of the wrist \cite{Sarac2020}, for different directions of skin deformation \cite{Sarac2022}, for different mapping strategies from multiple degree-of-freedom fingertip forces to single degree-of-freedom rendered forces \cite{Sarac2022_v2_wrist}, and different mapping strategies between index/thumb forces to dorsal/ventral wrist locations \cite{Palmer2022}.



 \vspace*{-.5\baselineskip}
\subsection{Motivation}

\subsubsection{Voice Coil Actuation}

Wrist-worn haptic devices for VR/AR applications are mostly based on DC/servo motors with cumbersome device sizes \cite{Sarac2022} or complex mechanical design \cite{Palmer2022, Pezent2022_wrist}. An alternative solution can be using voice coil actuation, which has been previously implemented for different haptic scenarios by rendering vibration feedback \cite{maeda_hapticaid_2016, gaudeni_presenting_2019} or single-bump force feedback (i.e., tapping) for social haptics \cite{culbertson2018social}, and for laparoscopic surgery \cite{tanaka2013tactile}. While custom-made voice coils can render better, more detectable, and more compelling single-bump force feedback as well \cite{Camardella:2022, Tanacar2023_fingertip}, there is no such technology implemented on the wrist. In this paper, we present a novel voice-\textbf{co}il based \textbf{wr}ist-worn \textbf{hap}tic device (CoWrHap) \cite{sen_design_2023}. \textbf{RQ1 --} Can minimalistic and trigger-based haptic rendering offer users a detailed and immersed stiffness exploration task?


\subsubsection{Hand Dominance}

While most exploration tasks are performed through dominant hands during visuo-haptic experiment protocols \cite{Palmer2022, Sarac2022, Pezent2022_wrist}, many realistic tasks (thus VR scenarios) involve using both hands \cite{Guo2021, Gruenewald2021}. Previous research shows no significant difference between using \textbf{dominant hands (DH)} in Fig. \ref{fig:conditions} (a, c) and \textbf{non-dominant hands (NDH)} in Fig. \ref{fig:conditions} (b, d) during object manipulation tasks \cite{Hu2018}. Yet, users' exploration capabilities while exploring the mechanical properties of virtual objects are to be discovered. \textbf{RQ2 --} Are there differences in users' performance, perception, or experience while exploring virtual objects using their DH compared to NDH?
\subsubsection{Hand-Wrist Congruence}

During exploration tasks, wrist-worn devices mostly render haptic information on users' dominant wrists as they interact with virtual tools using their dominant hands \cite{Palmer2022, Sarac2022, Pezent2022_wrist}. In this paper, this condition will be referred to as \textbf{hand-wrist congruence (H-WC)} in Fig. \ref{fig:conditions} (a, b). The literature also has successful applications of rendering the haptic feedback to the non-moving wrist \cite{gaudeni_presenting_2019} in the context of psychophysical research. In this paper, this condition will be referred to as \textbf{hand-wrist non-congruence (H-WNC)} in Fig. \ref{fig:conditions} (c, d). 
Considering that the sensory thresholds increase with movement \cite{Zhang2019}, there is no research comparing the impact of hand-wrist congruence -- especially during VR/AR interactions. We believe that H-WC might offer haptic cues to be more natural and compelling, while H-WNC might offer clearer and easier-to-interpret haptic cues to the user's skin since their hand movements might interfere with the human-machine interaction and the perceived forces. \\
\textbf{RQ3 --} Are there differences in users' performance, perception, or experience as they receive haptic feedback with H-WC compared to H-WNC?

 \vspace*{-.5\baselineskip}

\section{Voice-Coil based Wrist-Worn Haptic Device} \label{sec:wristworn}

\begin{figure}[h!]%
\centering
 \vspace*{-.75\baselineskip}
\includegraphics[width=0.4 \textwidth]{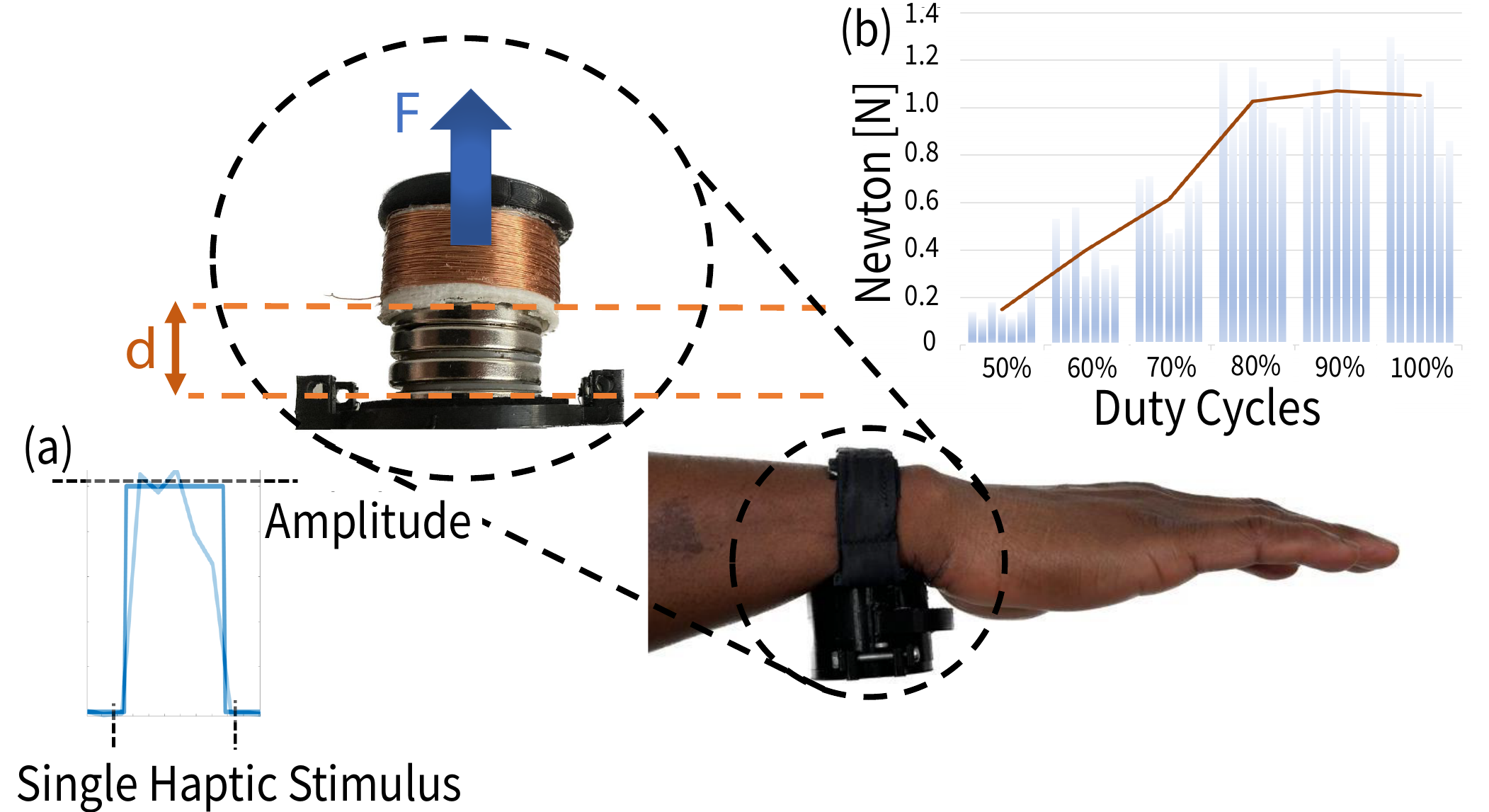}
 \vspace*{-.50\baselineskip}
 \caption{Custom-made CoWrHap representation. (a) It can render a single bump square signal. (b) The stimulus intensities can vary with an adjusted duty cycle between 50 \% and 100 \%, as indicated by force measurements.
 }
 \vspace*{-1\baselineskip}
 \label{fig:wristWornDevice}%
\end{figure}

CoWrHap in Fig. \ref{fig:wristWornDevice} consists of a permanent magnet and a voice coil built by wrapping a 0.25 mm copper wire around a cylindrical base with 35 mm in diameter and 20 mm in height – approximately 60 times \cite{sen_design_2023}. It weighs 160 grams. Changing the current levels passing through the wire creates a magnetic field inside the cylinder, which ultimately creates a displacement \textit{(d)} for the overall coil since the magnet is attached to the base. In the meantime, this deformation causes a corresponding interaction normal force on the user's skin \textit{(F)} as detailed in Fig. \ref{fig:wristWornDevice}. It is driven by an L293B motor driver and controlled with a Raspberry Pi Pico microcontroller. The current passing through the coil can be changed through levels of duty cycle percentages ranging from 0\% to 100\%. 

CoWrHap is driven and updated by a high-frequency Pulse-Width-Modulated (PWM) signal at 100 KHz. The duty cycle of the PWM signal is manipulated to change the amplitude of the haptic stimuli as needed while its duration is set to 0.5 seconds. Beyond 40\%, the duty cycle increases the current passing through the coil. As the duty cycle approaches 100\%, the current reaches a saturation point at approximately 930 mA, and its change diminishes. Additionally, the system requires a minimum input voltage of 10 Volts. 

CoWrHap was validated through preliminary studies based on force measurements and psychophysical discrimination abilities of users between different stimuli. Having a custom-made actuator is crucial in this context as it enables us to tailor design considerations, such as wire diameter and magnet choice. This flexibility allows for adjustments in the overall size of the device and the amount of achieved skin deformation. Consequently, the level of interaction forces can be fine-tuned based on the specific requirements of use cases. 


 \vspace*{-.5\baselineskip}
\subsection{Force Measurements}

A Force Sensitive Resistor (FSR) was placed between the user's skin and CoWrHap as it provides different levels of duty cycles. 
To minimize FSR's noisy behavior, repetitive force measurements were taken for each duty cycle between 50\% and 100\%. Although haptic feedback for 40\% duty cycle value was noticeable in the wrist, we did not observe FSR measurements for the duty cycle of 40\% and below. Fig. \ref{fig:wristWornDevice} (b) shows the 7 repetitive force measurements in blue bars between 50\% and 100\% duty cycle while the red, bold line represents their average. 
We also observed that after 80\%, measured forces saturate around 1 N, possibly due to the allowed current passing through the motor driver. For the rest of this paper, we will conduct our experiments within the linear range, where the duty cycles are set between 40\% to 80\%. 


 \vspace*{-.5\baselineskip}
\subsection{Preliminary Study on Detectability}

\begin{figure}[b!]%
\centering
\vspace*{-1\baselineskip}
\includegraphics[trim=0pt 0pt 0pt 30pt, clip, width=0.35\textwidth]{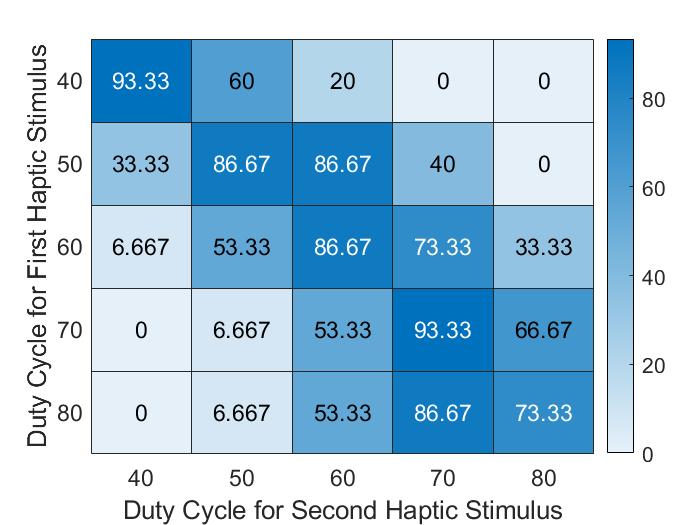}
\vspace*{-0.5\baselineskip}
\caption{Confusion matrix showing the user performance to discriminate two consecutive haptic stimuli with CoWrHap rendering randomized duty cycles.}
\label{fig:heatMap}
\end{figure}%

We then conducted a pilot study with 5 participants to validate their accuracy in detecting the differences between different stimuli -- similar to the method of constant stimuli. They were given two consecutive signals on their dominant wrist in a distinct order with various duty cycles in a randomized order three times (i.e., 40\% - 60\% three times and 60\% - 40\% another three times). Once the two consecutive signals are rendered, they are asked to verbally approve or disapprove that these two stimuli felt the same. Their accuracy for each pair was computed by averaging their correct answers and represented as a confusion matrix in Fig. \ref{fig:heatMap}. Ultimately, participants can discriminate stimuli with different duty cycles. 
Yet, their discrimination accuracy increases when (1) their duty cycle differences are greater and (2) at least one of the stimuli has a duty cycle lower than 70\%.


\section{Experiment Setup}

\begin{figure}[t!]%
\centering
\includegraphics[trim=0pt 0pt 0pt 0pt, clip, width=0.4\textwidth]{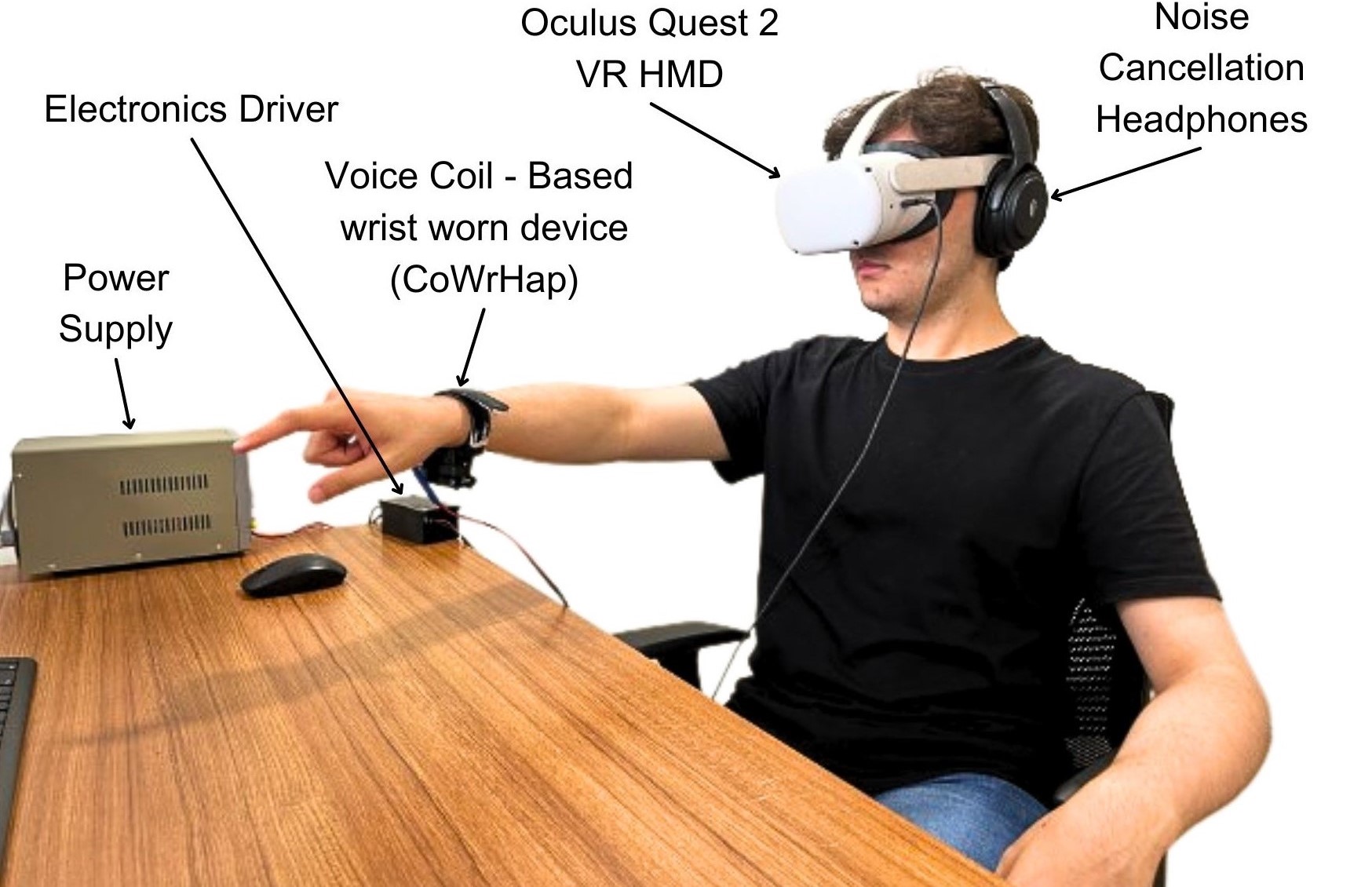}
 \vspace*{-.5\baselineskip}
 \caption{Experiment setup: The participant wears an Oculus headset to experience VR, CoWrHap to receive interaction-based haptic feedback, and noise-canceling headphones to minimize the environment and actuation noise.}
 \vspace*{-1\baselineskip}
 \label{fig:Setup}
\end{figure}

Fig. \ref{fig:Setup} shows the experiment setup: participants sit on a chair with arm supports and wear Oculus Quest 2 headset for (i) real-time hand tracking and (ii) visual representation of the VR environment. 
Once an interaction occurs between the tracked hand avatar and the virtual objects, CoWrHap renders haptic cues on the wrist based on its mechanical property. 
Noise-cancellation headphones with white noise minimize the environment and actuator noise.

 \vspace*{-.5\baselineskip}
\subsection{Virtual Environment}

We created a VR environment in Fig. \ref{fig:Scene} using Unity 2021.3.14f1 with a white desk, a hand avatar, and two identical-looking, red, rigid boxes to be displayed on Oculus Quest 2 headset with a refresh rate of 72 Hz. 
Hand Physics Toolkit \cite{jorgejgnz2020hptk} creates two hand avatars: \textit{(i)} a visible hand 
with limited motion when interacting with 
rigid objects and \textit{(ii)} an invisible 
hand always following users' movements. 
Without interaction, these two hands superimpose each other. With interaction, the visible hand is fixated on the object's surface while the haptic cues are rendered according to its stiffness. 

Triggered CoWrHap renders a bump-square signal as in Fig. \ref{fig:wristWornDevice} (a) whose intensity depends on the object's stiffness -- rather than being mapped to continuous force feedback based on Hooke's Law 
\cite{Sarac2022, Sarac2022_v2_wrist}. Relocating the haptic feedback and focusing on creating believable cues might also simplify the haptic feedback even further to single-bumped force feedback (i.e., tapping). This is particularly possible knowing that for visuo-haptic exploration environments, participants are more inclined to believe their visual perception than haptic \cite{Burns2006}. 


 \vspace*{-.5\baselineskip}
\subsection{Experiment Conditions}
To investigate our research questions, we designed a user study experiment with four main conditions through two factors: hand dominance ($2_{D} = $ DH, NDH) and hand-wrist congruence ($2_{C} = $ H-WC, H-WNC). In this paper, the term “congruence (C)" refers to the feedback being rendered on the wrist of the active interactive hand, while “non-congruence (NC)" refers to the feedback being rendered on the stationary wrist while the other hand interacts. These four experiment conditions can be summarized as in Fig. \ref{fig:conditions}:

\begin{itemize}
    \item \textbf{Dominant Hand-Wrist Congruence (DH-WC):} Participants interact with objects using their DH as haptic cues are rendered on their dominant wrist. 
    \item \textbf{Dominant Hand-Wrist Non-Congruence (DH-WNC):} They interact with objects using their DH as haptic cues are rendered on their non-dominant wrist. 
    \item \textbf{Non-Dominant Hand-Wrist Congruence (NDH-WC):} They interact with objects using their NDH as haptic cues are rendered on their dominant wrist.
    \item \textbf{Non-Dominant Hand-Wrist Non-Congruence (NDH-WNC):} They interact with objects using their NDH as haptic cues are rendered on their non-dominant wrist.
\end{itemize}

 \vspace*{-.5\baselineskip}
\subsection{Participants}
We conducted a user study with 28 participants (15 males and 13 females) with ages ranging between 18 and 26 years ($20.6 \pm 1.7$). The local Review Board approved the experimental protocol, and all
participants gave informed consent. 26 participants reported being right-handed, and 2 reported left-handed. 
19 participants had left-eye dominance, while 9 had right-eye dominance.  
Regarding their prior experiences with VR, 5 participants stated none, 11 stated one to two times, 6 stated three to four times, and 6 stated more than five times.

To keep the experiment time reasonably short with no potential fatigue for the participants, we conducted the experiment as a between-subjects design, where 14 participants perceived haptic feedback with H-WC and the rest with H-WNC. Regardless of the congruence, each participant interacted with the virtual tool using their DH and NDH in a randomized order.

\begin{figure*}[t!]%
\centering
\includegraphics[trim=0pt 0pt 0pt 0pt, clip, width=1\textwidth]{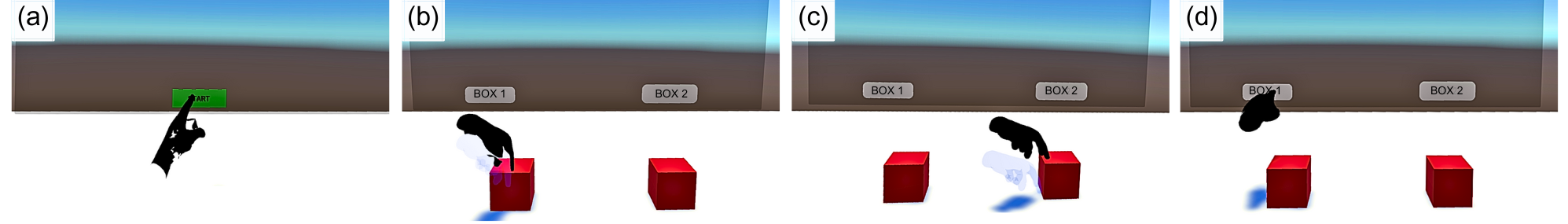}
 \vspace*{-2\baselineskip}
 \caption{Virtual task and the environment: the participant (a) selects the start button to initiate the trial, (b) explores Box 1 (on the left) by pushing from the top, (c) explores Box 2 (on the right) by pushing from the top, and (d) chooses the box which feels stiffer by clicking on the related button.}
 \vspace*{-1\baselineskip}
 \label{fig:Scene}%
\end{figure*}

 \vspace*{-.5\baselineskip}
\subsection{Experiment Protocol}


Fig. \ref{fig:Scene} shows the experiment flow. The participant initiates the trial by clicking on the “start'' button using the hand avatar. They interact with two identical-looking boxes with different stiffness values as many times as needed. Based on the haptic cues rendered through CoWrHap on their wrists, they choose the stiffer box by clicking on the button on top of it. 

Participants compared 7 different pairs 10 times each. During trials, one of the boxes had a constant (reference) stiffness, expressed with a duty cycle of 65\%. The other box had a varying stiffness, expressed with a duty cycle of 50, 55, 60, 65, 70, 75, or 80\%. 
We deliberately chose the range of duty cycles based on the results in Section \ref{sec:wristworn} and repeated the force measurements through 15 tests as seen in Table \ref{tab:Force Sensor Measurements}. We speculate that the noise-sensitive behavior of the FSR sensors significantly affects the standard deviation values. 

\begin{table}[h]
\centering
\vspace*{-.5\baselineskip}
\caption{Force measurements for the duty cycles used for the discrimination experiment below.}
\vspace*{-.5\baselineskip}
\label{tab:Force Sensor Measurements}
\resizebox{0.98\columnwidth}{!}{
\begin{tabular}{|c|c|c|c|c|c|c|c|}
  Duty Cycle(\%)    & 55& 60 & 65 & 70 & 75 & 80 & 100      \\
               \hline
Mean (N) & 0.336 & 0.455& 0.596 & 0.674 & 0.791& 0.871& 0.912 \\ \hline
Standard Errors & 0.039 & 0.029 & 0.029  & 0.028 & 0.034 & 0.035 & 0.031  \\ 
\end{tabular}}
\vspace*{-.25\baselineskip}
\end{table}

Ultimately, participants completed 140 comparison pairs to be discriminated -- 70 with their DH and 70 with their NDH. 
Even though the chosen feedback duration (i.e., 0.5 seconds) for each haptic cue did not cause CoWrHap to overheat, participants were given breaks for about 2 minutes after every 20 trials to guarantee an effective operation for the device. Finally, between the hand dominance conditions, they were given a longer break to switch the haptic device from their dominant wrist to their non-dominant wrist or vice versa. 
The overall experiment took around 30 minutes.

\section{Results}

We recorded and analyzed participants' performance in terms of the accuracy of their responses to the “stiffer'' box, how many times they tapped on the boxes, and how long it took them to make a decision. We also collected a post-experiment questionnaire about their overall experience. The data were processed on JMP and analyzed on SPSS 24. The figures show the mean in the graphs, and the error bars represent the standard error of the mean.  

\vspace*{-.5\baselineskip}
\subsection{Psychometric Curve}

\begin{figure}[h!]%
\centering
\includegraphics[trim=0pt 2pt 0pt 10pt, clip, width=0.45\textwidth]{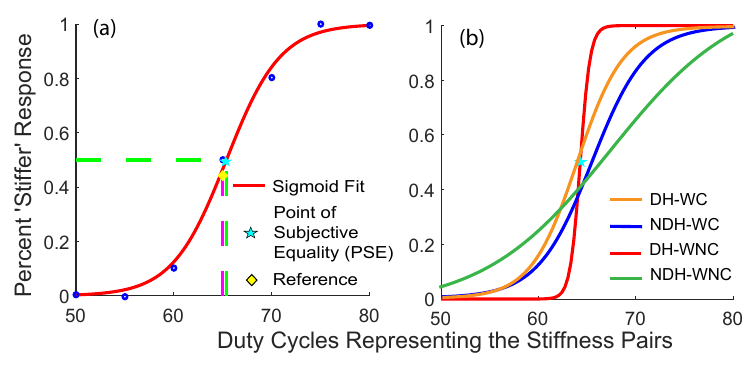}
 \vspace*{-1.0\baselineskip}
 \caption{Sample psychometric curves: (a) A Sigmoid fit and the PSE value are expressed through the collected data. (b) Obtained psychometric curves are in different forms for different experiment conditions.}
 \vspace*{-.5\baselineskip}
 \label{fig:All PSE}%
\end{figure}

Based on participants' discrimination accuracy, we formed a psychometric curve 
using the sigmoid function technique presented in a previous study \cite{Sarac2022} as in Fig. \ref{fig:All PSE}. Table \ref{tab:MeanValues} shows the mean and standard deviation of PSEs and JNDs. 

\begin{table}[h]
\centering
\vspace*{-1\baselineskip}
\caption{Mean Values of PSE and JND.}
\vspace*{-.5\baselineskip}
\label{tab:MeanValues}
\resizebox{0.98\columnwidth}{!}{
\begin{tabular}{|c|c|c|c|c|c|}
               & DH-WC             & NDH-WC            & DH-WNC            & NDH-WNC           \\
               \hline
PSE      & $66.69 \pm 2.56  $ & $67.59 \pm 2.51$ & $65.51 \pm 2.99 $ & $65.48 \pm 2.98$ \\ \hline
JND & $6.50 \pm 3.29$ & $8.69 \pm 6.10$ & $7.58\pm 2.78$ & $6.26 \pm 3.30$ \\ 
\end{tabular}}
\vspace*{-.5\baselineskip}
\end{table}

We analyzed the results using a two-way repeated-measures analysis of variance (RM-ANOVA) with dominance as within-subject and congruence as between-subject factors.

\begin{figure*}[t!]%
\centering
\includegraphics[trim=0pt 8pt 0pt 17pt, clip, width=1\textwidth]{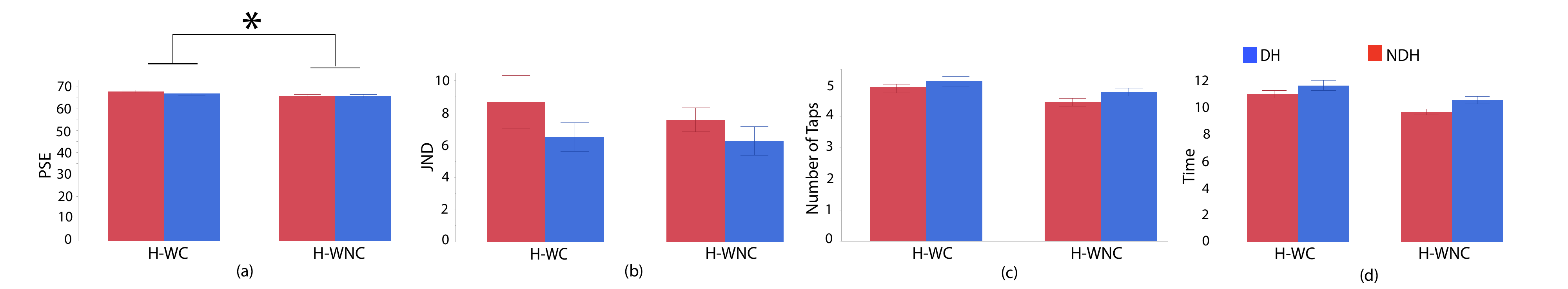}
 \vspace*{-1.5\baselineskip}
 \caption{Bar plots of results obtained from experiments based on congruence and hand dominance conditions in terms of average (a) PSEs, (b) JNDs, (c) the number of exploration taps, and (d) exploration time.}
 \vspace*{-.5\baselineskip}
 \label{fig:results}%
\end{figure*}

\textit{a) Point of Subjective Equality (PSEs)} relate to perceived sensation and stiffness intensity changes: closer PSEs to the reference indicate better discrimination performance. We observed that PSEs are statistically significantly higher with H-WC than H-WNC but not different between DH and NDH nor their interactions (Fig. \ref{fig:results} (a), Table \ref{tab:study2results}).

We also performed independent t-tests between the PSEs and the reference value (i.e., 65) for all four conditions. The results showed that the PSEs were statistically significantly different than the actual reference for H-WC conditions (DH-WC ($p = 0.020$) and NDH-WC ($p = 0.001$)) but not for H-WNC (DH-WNC ($p = 0.550$) and NDH-WNC ($p = 0.525$)). 

\textit{b) Just Noticeable Difference (JNDs)} relate to the smallest change perceived by participants: lower JNDs indicate better discrimination performance. 
We observed no statistically significant difference between dominance, congruence, or their interactions (Fig. \ref{fig:results} (b), Table \ref{tab:study2results}). 

\begin{table}[t!]
\centering
\caption{Statistical results from user study.}
\vspace*{-.5\baselineskip}
\label{tab:study2results}
\resizebox{\columnwidth}{!}{%
\begin{tabular}{|c|c|c|c|c|c|}
 &
  Hand Dominance &
  \begin{tabular}[c]{@{}c@{}} Hand-Wrist \\ Congruence \end{tabular}  & \begin{tabular}[c]{@{}c@{}} Dominance \\ X Congruence \end{tabular}
\\ \hline
\multicolumn{1}{|c|}{\begin{tabular}[c]{@{}c@{}} PSE \end{tabular}} &
  \begin{tabular}[c]{@{}c@{}} F(1, 26)=0.333, \\ p$=$0.569, $\eta^2$ = 0.013 
  \end{tabular} &
  \textbf{\begin{tabular}[c]{@{}c@{}}F(1, 26)=6.043, \\ p$=$0.021, $\eta^2=$ 0.189
  \end{tabular}} &
\begin{tabular}[c]{@{}c@{}}F(1, 26)=0.288, \\ p$=$0.596, $\eta^2=$ 0.011
\end{tabular} \\
  \hline
\multicolumn{1}{|c|}{\begin{tabular}[c]{@{}c@{}} JND \end{tabular}} &
  \begin{tabular}[c]{@{}c@{}}F(1, 26)=3.167, \\  p$=$0.087, $\eta^2$= 0.109
  \end{tabular} &
  \begin{tabular}[c]{@{}c@{}}F(1, 26)=0.327, \\ p$=$0.572, $\eta^2$=0.012
  \end{tabular} &
  \begin{tabular}[c]{@{}c@{}}F(1, 26)=0.194, \\ p$=$0.663, $\eta^2=$ 0.007
  \end{tabular}  \\ \hline
  \multicolumn{1}{|c|}{\begin{tabular}[c]{@{}c@{}} Number of \\ Taps on Boxes \end{tabular}} &
  \begin{tabular}[c]{@{}c@{}}F(1, 26)=0.873, \\ p$=$0.359,  $\eta^2$= 0.032
  \end{tabular} &
  \begin{tabular}[c]{@{}c@{}}F(1, 26)=0.199, \\ p$=$0.659,  $\eta^2$ = 0.008
  \end{tabular}  &
  \begin{tabular}[c]{@{}c@{}}F(1, 26)=0.070, \\ p$=$0.793, $\eta^2=$ 0.003
  \end{tabular}\\ \hline
\multicolumn{1}{|c|}{\begin{tabular}[c]{@{}c@{}}  Exploration \\ Time  \end{tabular}}  &
  \begin{tabular}[c]{@{}c@{}}F(1, 26)=1.383, \\ p$=$0.250,  $\eta^2$= 0.051
  \end{tabular} &
  \begin{tabular}[c]{@{}c@{}}F(1, 26)=0.324, \\ p$=$0.574,  $\eta^2$ = 0.012
  \end{tabular}  &
  \begin{tabular}[c]{@{}c@{}}F(1, 26)=0.028, \\ p$=$0.868, $\eta^2=$ 0.001
  \end{tabular} \\ 
\end{tabular}%
}
\vspace*{-1\baselineskip}
\end{table}


\vspace*{-.5\baselineskip}
\subsection{Discrimination Efforts}
We are also interested in discrimination efforts through the number of taps for the boxes and the exploration time for each comparison pair. 
We analyzed the results using a three-way RM-ANOVA with dominance and stiffness pairs as within-subject and congruence as a between-subject factors. 

\textit{a) The Number of Taps} relates to participants' level of confidence or confusion: lower numbers indicate a confident and clear decision. 
We observed no statistically significant difference among experiment conditions, stiffness pairs ($F_{1, \text{26}} = 1.594$, $p = 0.218$, $\eta^2 = 0.058$
), nor their interactions (Fig. \ref{fig:results} (c), Table \ref{tab:study2results}). 

\textit{b) Exploration Time} relates to participants' level of confidence or confusion: lower time indicates a more confident decision. 
From our results, we deduce that there is no statistically significant difference among experiment conditions, the stiffness pairs ($F_{1, \text{26}} = 1.987$, $p = 0.171$, $\eta^2 = 0.071$
), nor their interactions (Fig. \ref{fig:results} (d), Table \ref{tab:study2results}). 

\vspace*{-.5\baselineskip}
\subsection{Subjective Questionnaire}



Participants completed a post-questionnaire on their experience and preferences. 20 participants preferred using their dominant hand and 8 their non-dominant hand. They were also asked to rate various user experience concepts on a 7-point Likert scale, as shown in Table \ref{tab:SubjectiveTable}. Participants reported the task to be easier for \textit{(i)} H-WC group than H-WNC and \textit{(ii)} using DH than NDH. 
Both participant groups rated higher rates of pleasantness for haptic cues using DH than NDH. Finally, asked about the pleasantness of the discrimination task, \textit{(i)} both groups rated DH higher than NDH, and \textit{(ii)} their difference was higher with H-WNC than H-WC.

\begin{table}[h]
\centering
\vspace*{-1\baselineskip}
\caption{Post experiment questionnaire results.}
\vspace*{-.5\baselineskip}
\label{tab:SubjectiveTable}
\resizebox{0.98\columnwidth}{!}{
\begin{tabular}{|c|c|c|c|c|c|}
               & \multicolumn{2}{c|}{\textbf{H-WC}} & \multicolumn{2}{c|}{\textbf{H-WNC}}\\
               \hline
Physical Fatigue & \multicolumn{2}{c|}{$4.4 \pm 1.3$} & \multicolumn{2}{c|}{$3.1 \pm 1.8$} \\ \hline
Mental Fatigue & \multicolumn{2}{c|}{$3.4 \pm 2.3$} & \multicolumn{2}{c|}{$2.0 \pm 1.4$} \\ \hline \hline
               & \textbf{DH-WC}             & \textbf{NDH-WC }           & \textbf{DH-WNC }           & \textbf{NDH-WNC}           \\
               \hline
Task Ease      & $5.0 \pm 1.4$ & $4.3 \pm 1.6$ & $4.4\pm 1.6$ & $3.5 \pm 1.8$ \\ \hline
Haptic Feeling & $5.5 \pm 1.8$ & $4.6 \pm 2.3$ & $5.4\pm 1.4$ & $4.6 \pm 2.1$ \\ \hline
Discrimination & $5.1 \pm1.7$ & $4.6 \pm 1.8$ & $5.9 \pm 1.7$ & $4.0 \pm 2.0$

\end{tabular}}
\vspace*{-.5\baselineskip}
\end{table}

Participants also rated having perceived fairly believable interactions and a sense of touch ($4.61 \pm 1.50$). Finally, when asked \textit{While making your selections, which strategy did you use? What did you focus on?} in an open question form, some participants wrote ‘‘I felt the soft box more, so it was easy to know the hard box" or ‘‘I focused on the first box that I touched; in that way, I could differentiate the stiffness values'' -- supporting their reported rate values.

\section{Discussions and Future Works}

In this paper, we designed a novel, custom-made device, CoWrHap, using a wire and a magnet. CoWrHap offers distinct advantages. It is cost-effective compared to commercial force feedback haptic devices and can be easily customized specifically for responsive touch in VR tasks based on user feedback or performance. Our quantitative and subjective results indicate that participants can comfortably notice the differences among haptic stimuli from 40\% to 100\% duty cycle. 
The results of our discrimination experiment indicate that \textit{participants could successfully discriminate objects with different stiffness levels in VR while using single-bump force feedback generated by voice-coil actuation, following \textbf{RQ1}.} 

All participants were asked to perform the discrimination tasks using their dominant and non-dominant hands, and we analyzed the impact of hand dominance in terms of quantitative measures (e.g., PSEs, JNDs, number of taps, exploration time) and subjective comments. While we expected participants to show better discrimination performance using their dominant hands, our results indicated no statistical difference. 
This could indicate that participants could perceive mechanical properties 
confidently since they received haptic feedback through CoWrHap in a meaningful, believable, and noticeable manner, shadowing the unnaturalness of interactions with their non-dominant hand. We speculate that this is due to the likelihood and success of humans using both hands in daily activities more than expected. Such unintentional habits become more apparent when participants navigate a completely new virtual environment. Yet, the subjective comments yield that participants had a better experience while performing the task using their DH in terms of ease of task, enjoyment, and perceived fatigue. Thus, regarding \textit{\textbf{RQ2 --} virtual discrimination tasks using dominant or non-dominant hands show no perceptual or performance differences, even though participants felt more comfortable using their DH.}

Participants were randomly divided into two groups: one to receive haptic feedback on the wrist in congruence with the hand and one in non-congruence. Our results 
show better psychophysical performance with H-WNC in terms of PSEs only -- although the H-WC group reported a better user experience through subjective comments. 
We speculate that the participants found the H-WNC conditions physically and mentally more tiring – while their higher sensitivity to the lack of movement (also supported by the literature \cite{Zhang2019}) somewhat balanced the overall discrimination performance. In summary, regarding \textit{\textbf{RQ3 --} congruence conditions indicate differences in terms of user performance and experience: H-WNC offers significantly better performance, yet H-WC offers a more natural user experience regardless of hand dominance.} 
We speculate that extending the experiment to more participants might yield further advantages of H-WNC over H-WC.  

\vspace*{-.5\baselineskip}
\subsection{Limitations and Future Work}

For ergonomics, participants agreed that CoWrHap is comfortable to wear and can render pleasant haptic feedback -- except one suggested the device could be smaller and lighter for better comfort in the future.CoWrHap's custom-made actuation unit allows for size adjustments by altering heights, diameters, rolls, and wire type. In the meantime, the force measurements of CoWrHap were observed to saturate around 1N and after 80\% duty cycle for the PWM signals. Ultimately, with the chosen electronics equipment, CoWrHap cannot be operated to its fully potential, while still providing comfortable, meaningful, and distinguishable stimuli. Future investigations will explore minimizing its size for better comfort, which can use PWM signals up to 100\% duty cycle without sacrificing the levels of interaction forces. 

We acknowledge the inconsistencies in the force responses of CoWrHap due to the noise-sensitive behavior of the FSR sensor. In the future, we will characterize the forces using more capable load cells and force sensors and optimize the size of CoWrHap without changing the output forces. 

Regarding the discrimination task, rendering single-bumped force feedback is not a realistic way of implementing stiffness exploration scenarios in a VR setting. Still, our results indicate the potential for using voice-coil actuation in such a detailed and immersed stiffness exploration task -- while also bringing a few disadvantages. \textit{(i)} Their performance over continuous haptic feedback is still unknown -- providing the force feedback with a single-bump instead of a continuous cue might change the user behavior and ultimate performance during stiffness exploration tasks. \textit{(ii)} These results might not extend to different exploration tasks that require the acquisition of continuous haptic knowledge on the user side, like friction exploration. Finally, our  study focused on single-handed interaction with virtual boxes. In the future, we would like to first develop 
and utilize haptic devices with continuous force feedback (e.g., linear or servo motors) and compare it to using CoWrHap and then explore the potential for valuable insights in scenarios involving bi-lateral use.


\section{Conclusions}

This paper presents a custom-made voice-coil actuated wrist-worn haptic device (CoWrHap) to render single-bump force feedback in response to VR interactions. We also investigated how hand dominance and the congruence between the hand and the wrist influence their task performance, perception, and preferences during a discrimination experiment. 
Our findings demonstrated that participants' discrimination performance was unaffected by hand dominance even though they reported preferring using their dominant hand. Yet, their performance was observed to be better with H-WNC (i.e., when the haptic feedback is rendered on a stationary wrist rather than the moving one), even though they preferred congruence in terms of user experience.

Notably, the low-cost and easily customizable nature of voice coils positions them as a promising alternative among other actuators, enhancing the adaptability and accessibility of our custom-made devices. Ultimately, the insights gained from these studies are poised to guide future designers and hapticists in seamlessly integrating haptics into augmented reality interactions. In the future, our research will evolve further around characterizing and optimizing CoWrHap and compare it to other actuation technologies during different discrimination tasks and virtual interaction scenarios. 


\ifCLASSOPTIONcaptionsoff
  \newpage
\fi



\bibliographystyle{IEEEtran}
\bibliography{References}
%

%








\end{document}